\documentclass[preprint,12pt,number]{elsarticle}

\usepackage{graphicx}
\usepackage{epsfig}
\usepackage{amssymb}

 \journal{Chemical Physics}

\begin{document}

\begin{frontmatter}



\title{\textbf{The Complex Energy Spectrum of Isomeric Reactions}}


\author[a]{A.~Ugulava}
\author [a]{Z.~Toklikishvili}
\author[a]{S.~Chkhaidze\corref{*}}
\ead{s.chkhaidze@tsu.ge}
\author[b,c]{L.~Chotorlishvili}
\author [a]{R.~Abramishvili}
\cortext[a]{Corresponding author}
\address[a]{I.Javakhishvili Tbilisi State University,
I.Chavchavadze av., 3, Tbilisi 0128, Georgia}
\address[b]{Institut f\"ur Physik, Martin-Luther Universit\"at
Halle-Wittenberg, Heinrich-Damerow-Str.4, 06120 Halle, Germany}
\address[c]{Institut f\"ur Theoretische Physik Universit\"at
Heidelberg Philosophenweg 19, D-69120 Heidelberg}
\begin{abstract}

The internal motion in a molecule, in which isomerization
processes occur, is characterized by two essentially different
modes of motion - oscillatory and rotational. The quantum equation
of motion which describes an isomerization process is reduced to
the Mathieu-Hill equation. As is known, this equation is able to
describe both modes. In the paper, it is shown that the chaotic
region of the energy spectrum characterizing an isomerization
process corresponds to the region where two modes of motion are
merged together.
\end{abstract}

\end{frontmatter}


\section{Introduction}
\label{1}

Dynamic stochasticity
\cite{Lichtenberg-83,Sagdeev-88,Alligood-96}\ occurs in the region
of parameter values, where topologically different trajectories
adjoin each other. Trajectories near the boundary are highly
sensitive to small disturbances. For small disturbances these
trajectories take a very complicated shape, which is one of the
manifestations of stochasticity.

An elementary example of such systems is a pendulum that may
perform two kinds of motion - oscillatory and rotational. This
explains in the main a growing interest in physical processes in
which motion reduces to pendulum type motions.

A quantum analogue of dynamic stochasticity - quantum chaos
\cite{Stockmann-93, Haake-01}\ occurs in the same region of
parameter values where dynamic stochasticity does. In that case,
the quantum states of two different symmetries adjoin each other.
A highly complicated process takes place in a neighborhood of the
boundary - the energy spectrum corresponding to the symmetry of
one kind transforms to the energy spectrum of the other kind,
which, in turn, demands the removal of degeneration corresponding
to the symmetry of one kind and the appearance of degeneration
corresponding to the symmetry of second kind. In the energy
spectrum graph this process is represented by the set of branch
points and merged energy levels. In passing through these points,
the system may lose information on the initial conditions and pass
from the pure state to the mixed one. The mixed state thus formed
will be the state expressing quantum chaos. From that moment of
time, the state of the system is not any longer described by a
wave function and it becomes necessary to describe it by a density
matrix \cite{Chotorlishvili-10,Ugulava-04,Ugulava-05}.

Frequently, the objects of quantum chaos realization are
polyatomic molecules
\cite{Essers-82,Arranz-98,Eryomin-2000,Ugulava-07}. For example,
quantum chaos may take place in a polyatomic molecule, one of the
generalized coordinates of which performs torsional oscillations
which, with an increase of the amplitude, transform to rotational
motion \cite{Ugulava-07}. Quantum chaos is a characteristic
feature of such a transitional region.

From the standpoint of quantum chaos it is of special interest to
investigate the so-called floppy molecules which have a capacity
to isomerization \cite{Essers-82,Arranz-98}. An example of such
molecules is lithium cyanide (LiNC). This molecule contains a firm
fragment\ $C\equiv N$\ with a triple bond and a relatively light
lithium atom (Li) which is able to perform angular oscillations
relative to the molecule axis. When the oscillation amplitude
attains the threshold value, the lithium atom detaches itself from
the nitrogen atom and passes over to the carbon atom (Fig.1a). At
higher energies, it again returns to the nitrogen atom and there
begins the alternating jump-over process, i.e., the lithium atom
begins to rotate about the fragment\ $C\equiv N$.

The energy spectrum of lithium atom motion was studied by the
numerical methods in \cite{Essers-82,Arranz-98}. It was shown that
the energy spectrum consists of equidistant and nonequidistant
regions. The energy levels arranged in a nonequidistant manner
correspond to lithium atom states at which the oscillatory mode of
motion transforms to the rotational one.

The aim of the present paper is to study the energy spectrum of
lithium atom motion relative to the firm fragment\ $C\equiv N$. As
different from the papers \cite{Essers-82,Arranz-98},\ which are
based mainly on the numerical methods, our approach to the problem
is analytical: using numerical data we approximate the potential
energy of the lithium atom and study the energy spectrum by the
analysis of the Schr\"odinger equation.

\section{\textbf{An Isomerization Potential}}
\label{2}

The lithium atom motion relative to the firm fragment\ $C\equiv
N$\ can be described by means of two coordinates \emph{R} and
$\theta$\ only (Fig.1b). $\theta=0^0$ and $\theta=180^0$
correspond to the equilibrium states of Li--CN and NC--Li,
respectively. $\emph{R}$\ is the distance between the lithium atom
and the centre of mass of the fragment $C\equiv N$.

\begin{figure}[!h]
\centering -\includegraphics [width=10cm]{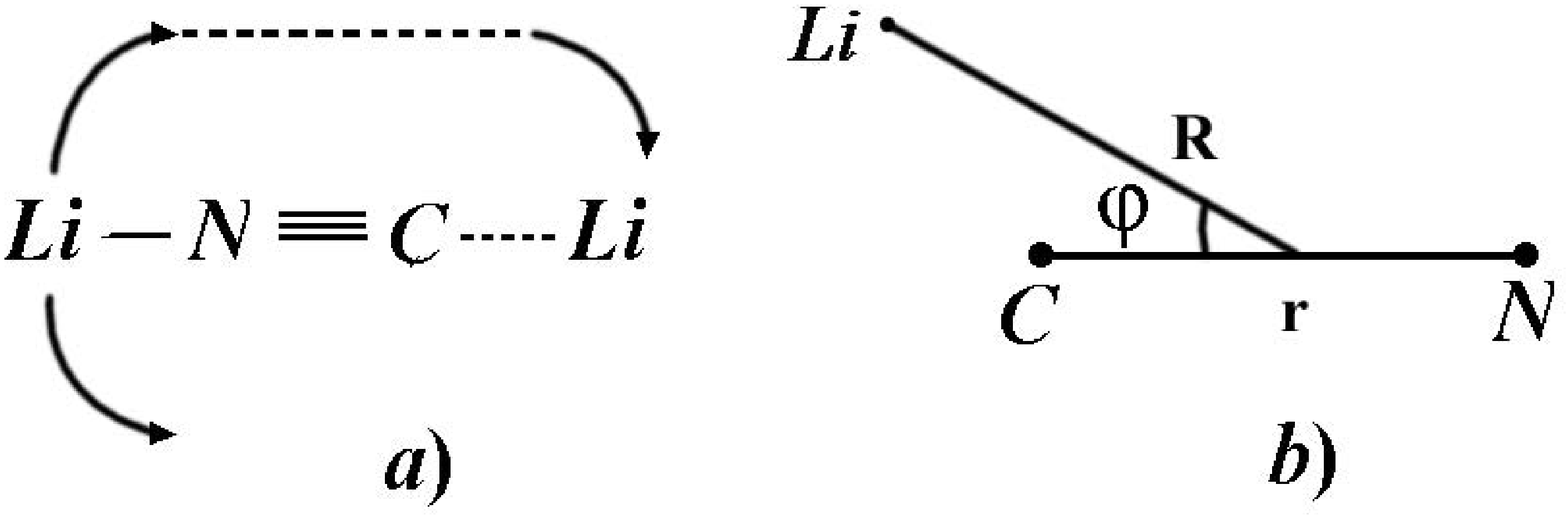} \caption{The
izomerization reaction in lithium cyanide: $\emph{a}$) lithium
atom motion relative to the firm fragment; $\emph{b}$) the
coordinates $\emph{R}$ and $\theta$ describing the lithium
atom.}\label{fig:1}
\end{figure}

The potential energy \emph{V} of lithium atom motion relative to
the firm fragment $C\equiv N$ was studied in \cite{Essers-82}
(Fig.2). It was shown that the potential energy has two different
minima for $\theta=0^0$\ and $\theta=180^0$. The energy distance
between these minima is $\emph{V}(180^0)-\emph{V}(0)\approx6.21$
kcal/mol,while the distance from the minimum at $\theta=0^0$ to
the maximum is 3.42 kcal/mol. Totally, the isomerization energy,
i.e. the distance from $\emph{V}(180^0)$ to $\emph{V}_{max}$ is
$\approx9.63$ kcal/mol (Fig.2).

\begin{figure}[!h]
\centering
\includegraphics[width=8cm]{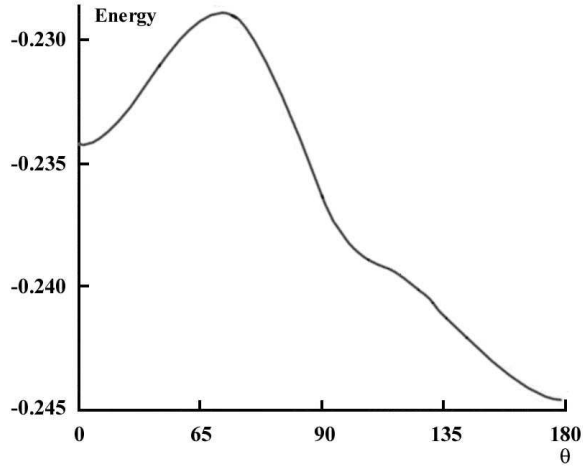}
\caption{The plot of the potential energy $\emph{V}(\theta)$
obtained by a numerical method \cite{Essers-82}. The angle
$\theta$ is laid off along the abscissa axis, and the potential
energy in atom units $(1\ am.u.=627.52\ kcal/mol, 1\
kcal/mol=7\cdot10^{-21}J)$ is laid off along the ordinate axis.
}\label{fig:2}
\end{figure}

It can be easily verified that the potential energy of
isomerization is approximated by the formula
$$V(\theta)=C+\alpha_1\cos\theta-\alpha_2cos^2\theta,$$
where $\alpha_{1}<2\alpha_{2}$. Indeed, the points of a minimum of
the potential energy $V(\theta)$ are $\theta=\pi{n}$, while the
points of a maximum
$\theta_\pm=\pm\arccos\frac{\alpha_1}{2\alpha_2}+2\pi{n}$. If
$\alpha_1\rightarrow2\alpha_2$, then
$\theta_+\rightarrow\theta_-\rightarrow2\pi{n}$. This means that
minima vanish at points $2\pi{n}$, while maxima adjacent to them
merge. Therefore $\alpha_1<2\alpha_2$ is the condition of
existence of two pits with different depth values.

We begin the reading of the potential energy $\emph{V}(\theta)$
from the largest minimum $\emph{V}[(2n+1)\pi]$. Then
$C=\alpha_1+\alpha_2$, and for defining the numerical values of
$\alpha_1$ and $\alpha_2$ we obtain the equations
$\emph{V}(\theta)=2\alpha_1$, $\emph{V}(\pi)=0$,
$\emph{V}_{max}=\emph{V}(\theta_\pm)=\frac{(2\alpha_2+\alpha_1)^2}{4\alpha_2^2},$
from which for the above-mentioned numerical values we obtain
$\alpha=3.245$ kcal/mol and $\alpha_2=6.243$ kcal/mol.

The plot of the approximated potential energy is shown in Fig.3.
It should be noted that the existence of potential minima of
different heights is typical of isomerization processes where the
firm fragment is nonsymmetric \cite{Herzberg-45}.

\begin{figure}[!h]
\centering \includegraphics[width=8cm]{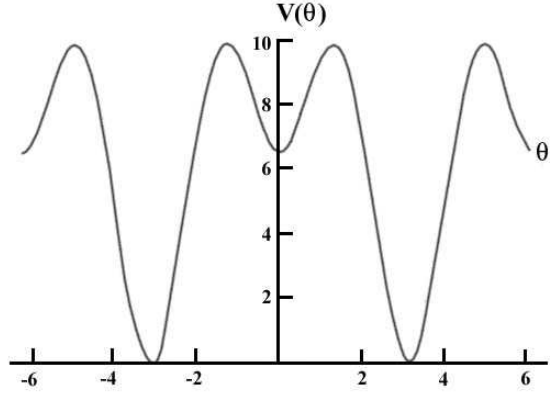} \caption{The
potential energy of the lithium atom representing the
isomerization process. The graph is constructed for the potential
energy
$V(\theta)=\alpha_2+\alpha_1+\alpha_1\cos\theta-\alpha_2\cos^2\theta,$
when $\alpha_{1}<2\alpha_{2}$.}\label{fig:3}
\end{figure}
A Hamiltonian function describing the lithium atom motion relative
to the firm fragment $C\equiv N$ has the form
\begin{equation} H=\frac{P^2_\theta}{2I}+V(\theta),
\end{equation}
where $I=M_1R^2+M_2r^2$ is the moment of inertia, and $P_\theta$
is the generalized pulse corresponding to the cyclic variable
$\theta$.

\begin{equation}
\frac{1}{M_1}=\frac{1}{m_{Li}}+\frac{1}{m_N+m_C},
\qquad\frac{1}{M_2}=\frac{1}{m_N}+\frac{1}{m_C},
\end{equation}

\begin{equation}
V(\theta)=\alpha_2+\alpha_1+\alpha_1\cos\theta-\alpha_2\cos^2\theta.
\end{equation}
$m_{Li}$, $m_N$ and $m_C$ are the masses of lithium, nitrogen and
carbon atoms. The expression (1) does not take into account the
radial motion of the lithium atom, which is assumed to be at some
average distance from the fragment $C\equiv N$.

\section{The Classical Consideration. The Phase Portrait}
\label{3}

Let us first consider the lithium atom motion relative to the firm
fragment $C\equiv N$ in classical terms. For the sake of
simplicity, we will study this motion for energy $E>2\alpha$. As
will be seen later, these energy values include the oscillation
and rotation modes which we are interested in.

To obtain the motion equation, we write the energy integral

\begin{equation}
\frac{I\dot{\theta}^2}{2}+\alpha_{1}+\alpha_{2}+\alpha_{1}\cos\theta-\alpha_2\cos^2\theta=E.
\end{equation}

Using this integral of motion we can obtain a period of the
lithium atom rotation about the fragment $C\equiv N$

$$T_+=2\sqrt{\frac{I}{2}}\int\limits_{0}^{\pi}\frac{d\theta}{\sqrt{E-U(\theta)}}=2\sqrt{\frac{I}{2}}\sum_\pm
F(\gamma_\pm,r),$$ where
$$\gamma_{\pm}=\arcsin\left(\left(1\pm\frac{\alpha_1}{2\alpha_2}\right)\sqrt{\frac{a^2+1}{a^2+\left(1\pm\frac{\alpha_1}{2\alpha_2}\right)^2}}\right),
\qquad
a^2=\frac{E}{\alpha_2}-\left(1+\frac{\alpha_1}{2\alpha_2}\right)^2,$$
$$r=\sqrt\frac{(1-b)(c+1)}{(1-c)(b+1)},
\qquad
b=\frac{\alpha_1}{2\alpha_2}+\sqrt{\left(1+\frac{\alpha_1}{2\alpha_2}\right)^2-\frac{E}{\alpha_2}},$$
$$c=\frac{\alpha_1}{2\alpha_2}-\sqrt{\left(1+\frac{\alpha_1}{2\alpha_2}\right)^2-\frac{E}{\alpha_2}}.$$
Here
$$F(\alpha,k)=\int_{0}^{\alpha}\frac{d\varphi}{\sqrt{1-k^2\sin^2\varphi}}$$
is a eliptic function of first order.

When
$E\rightarrow+E_c=\frac{\left(2\alpha_2+\alpha_1\right)^2}{4\alpha_2},$
in the limit from above we have $a\rightarrow0$,\
$\gamma_\pm\rightarrow\frac{\pi}{2}$,\ $r\rightarrow1$, and the
rotation period
$$T_+\rightarrow+T_c=4\sqrt{\frac{I}{2\alpha_2}}F\left(\frac{\pi}{2},r\rightarrow1\right)=4\sqrt{\frac{I}{2\alpha_2}}\ln\frac{4}{\sqrt{1-r^2}}\rightarrow\infty.$$
Therefore as $E\rightarrow+E_c$ the rotation period
logarithmically tends to infinity. Analogosously, we can calculate
the oscillatory motion period $T_-$. These calculations are not
given here. We only note the oscillatory motion period, too,
diverges logarithmically when $E\rightarrow-E_c$,\ i.e., if
$E\rightarrow-E_c$, then $T_-\rightarrow-T_c\rightarrow\infty$.

As a rule, the existence of a logarithmically diverging period
indicates the presence of two different modes of motion and the
existence of a boundary trajectory - separatrix between them. This
means that the separatrix appears to be obtained at energies
$E=-E_c$.

Using the expression (4) we can construct the phase portrait on
the plane $(\dot\theta,\theta)$. As seen from Fig.4, the phase
portrait consists of closed (elliptic) trajectories, which
correspond to the oscillatory process, and wave trajectories
describing rotational motion. The alternation of elliptic
trajectories of two different dimensions on the $\theta$-axis is
caused by the presence of two different minima of potential
energy. Thus we see that lithium atoms may have two types of
motion - oscillatory when
$E<\frac{(2\alpha_2+\alpha_1)^2}{4\alpha_2}$ and rotatory when
$E>\frac{(2\alpha_2+\alpha_1)^2}{4\alpha_2}$. As seen from the
figure these two types of motion are divided by the separatrix.
\begin{figure}[!h]
\centering -\includegraphics [width=6cm]{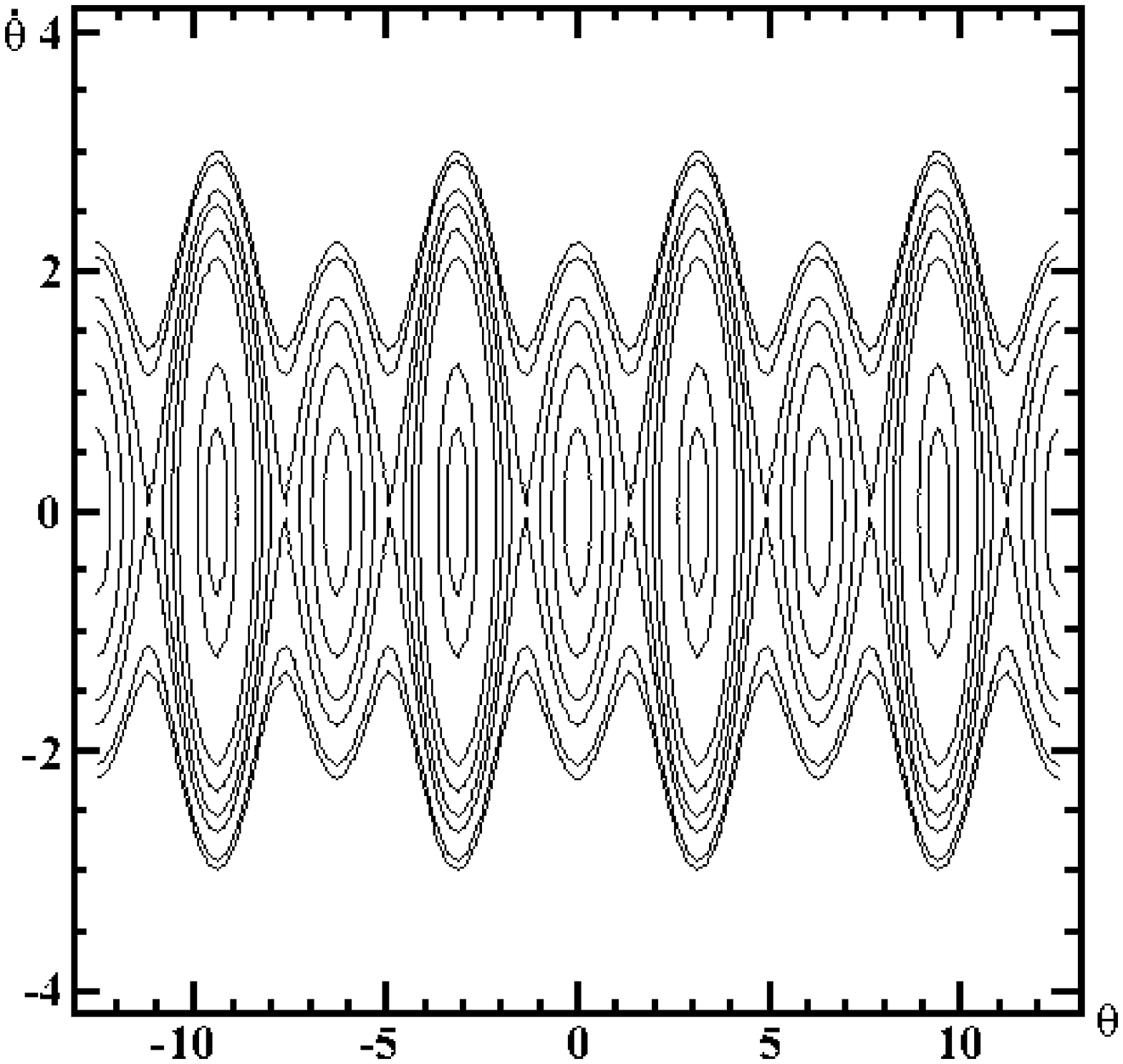} \caption{The
phase portrait corresponding to an isomerization process,
constructed by means of the equation (4).}\label{fig:4}
\end{figure}

\section{Quantum-Mechanical Consideration of Motion. The Mathieu-Hill- Schr\"odinger Equation.}
\label{4}

Let us now consider the motion of the lithium atom relative to the
fragment $C\equiv N$ from the standpoint of quantum mechanics. We
replace the Hamiltonian function (1) by the Hamilton operator

\begin{equation}
\hat{H}=\frac{\hat{p}^2}{2I}+V(\theta),
\end{equation}
where $\hat{p}=-i\hbar\frac{\partial}{\partial\theta}$ is a pulse
operator. Then the Schr\"odinger equation takes the form

\begin{equation}
-\frac{\hbar^2}{2I}\frac{\partial^2\psi}{\partial\theta^2}+(\alpha_1+\alpha_2+\alpha_1\cos\theta-\alpha_2\cos^2\theta)\psi=\varepsilon\psi.
\end{equation}
We make the replacement $\theta\rightarrow2\theta$ and introduce
the notation
$$E=\frac{8I}{\hbar^2}\left(\epsilon-\alpha_1-\alpha_2+\frac{\alpha_2}{2}\right),
\qquad q_1=\frac{8I}{\hbar^2}\alpha_1, \qquad
q_2=-\frac{8I}{2\hbar^2}\alpha_2.$$ Then, by simple
transformations, from the equation (6) we obtain
\begin{equation}
\frac{d^2\psi}{d\theta^2}+\left[E-2\lambda(q_1\cos2\theta+q_2\cos4\theta\right)]\psi=0.
\end{equation}
For the sake of simplicity, we restrict our consideration to the
investigation of the dependence of energy levels on one (and not
on two) parameter and to this end we introduce the parameter
$\lambda$ into (7). This relationship is studied assuming $q_1$
and $q_2$ to be given values. The energy spectrum of the initial
problem is obtained at $\lambda=0.5$. The Schr\"odinger equation
takes the form known as the Hill equation which in turn implies
the Mathieu equation as $q_1\rightarrow 0$. Using the known
methods \cite{McLachlan-47},\ solutions of this equation are
sought for in the form of Fourier series
\begin{eqnarray}
Ce_{2n}(z;\lambda;q_1,q_2)&=&\sum_{r=0}^{\infty}A_{2r}^{(2n)}(\lambda)\cos{2rz},\nonumber\\
Ce_{2n+1}(z;\lambda;q_1,q_2)&=&\sum_{r=0}^{\infty}A_{2r+1}^{(2n+1)}(\lambda)\cos(2r+1)z.\nonumber\\
Se_{2n+1}(z;\lambda;q_1,q_2)&=&\sum_{r=0}^{\infty}B_{2r+1}^{(2n+1)}(\lambda)\sin(2r+1)z,\nonumber\\
Se_{2n+2}(z;\lambda;q_1,q_2)&=&\sum_{r=0}^{\infty}B_{2r+2}^{(2n+2)}(\lambda)\sin(2r+2)z.
\end{eqnarray}
Substituting these values into the equation (7) and equating the
coefficients of the same harmonic, we obtain recurrent relations
for the coefficients $A_{2r}$,\ $A_{2r+1}$,\ $B_{2r}$,\
$B_{2r+1}$, (their upper indexes are omitted). For example, for
the first series of (8) these relations have the form:
\begin{eqnarray}
r&=&0\qquad aA_0-\lambda{q_1A_2}-\lambda{q_2A_4}=0, \nonumber\\
r&=&1\qquad(a-4)A_2-2\lambda{q_1A_0}-\lambda{q_1A_4}-\lambda{q_2A_2}-\lambda{q_2A_6}=0,\nonumber\\
r&=&2\qquad(a-16)A_4-\lambda{q_1A_2}-\lambda{q_1A_6-2}\lambda{q_2A_0}-\lambda{q_2A_8}=0,\nonumber\\
r&=&3\qquad(a-36)A_6-\lambda{q_1A_4}-\lambda{q_1A_8}-\lambda{q_2A_2}-\lambda{q_2A_{10}}=0,\nonumber\\
r&\geq&4\qquad(a-4r^2)A_{2r}-\lambda{q_1A_{2(r-1)}}-\lambda{q_1A_{2(r+1)}}-\lambda{q_2A_{2(r-2)}}-\nonumber\\
&&-\lambda{q_2A_{2(r+2)}}=0.
\end{eqnarray}
This is an infinite system of equations that establishes a
relation between the coefficients $A_{2r}$. In order that this
system would have a nontrivial solution it is necessary that
$Det(A_{ij})=0$, where
\begin{eqnarray}
A_{ij}&=&\left[a-4(i-1)^2\right]\delta_{ij}-\lambda{q_1\delta_{i+1;j}}-\lambda{q_2\delta_{i+2;j}}-\lambda{q_1\delta_{i;j+1}}-\lambda{q_2\delta_{i;j+2}}-\nonumber\\
&&-\lambda{q_1\delta_{i;2}\delta_{j;1}}-\lambda{q_2\delta_{i;3}\delta_{j;1}}-\lambda{q_2\delta_{i;2}\delta_{j;2}},\
i=1\ldots\infty,\ j=1\ldots\infty.
\end{eqnarray}
are elements of the matrix composed of the coefficients of the
system of equations (9) and written by means of the Kronecker
symbol. Analogously, we obtain recurrent relations for the
coefficients $A_{2r+1}$,\ $B_{2r}$ and $B_{2r+1}$, matrix
determinants and characteristic equations, the numerical solution
of which gives the energy spectrum.

\section{A Quantitative Estimate of an Energy Spectrum}
\label{5}

To obtain a quantitative estimate of the energy spectrum, we have
to calculate first the numerical values of the parameters  $q_1$
and $q_2$ contained in the Mathieu-Hill equation (7).

Applying the obtained values of $\alpha_1$ and $\alpha_2$, taking
into consideration the carbon, nitrogen and lithium mass values
($m_C=19.92\cdot10^{-27}kg,$\ $m_N=23.24\cdot10^{-27}kg,$\
$m_{Li}=11.62\cdot10^{-27}kg$) and also the values of the linear
parameters $\emph{r}$ and $\emph{R}$ from \cite{Essers-82}\
($r=2.186a_0=11.57\cdot10^{-11}m,$ \
$R=4.05a_0=21.43\cdot10^{-11}m,$), we obtain, by means of the
equations (2),$q_1=\frac{8I}{\hbar^2}\alpha_1\approx9300$ and
$q_2=-\frac{8I}{\hbar^2}\alpha_2\approx-8942$. Using the obtained
values of $q_1$ and $q_2$ and the numerical solutions of the
characteristic equation, we can construct the energy spectrum as a
function of Mathieu characteristics $E(\lambda)$ (Fig.5).

As seen from Fig.5a, for low energies we have levels arranged
equidistantly, which corresponds to oscillations of the lithium
atom near the nitrogen molecule, i.e., to oscillatory motion in a
deep ($\theta=0^0$) pit (Fig.3). When the lithium atom energy is
$\varepsilon\geq6\mbox{\ kcal/mol}$, the alternating transition
from one pit to the other and vice versa takes place. In Fig.5 to
this transition there correspond levels arranged nonequidistantly.
The spectrum obtained by us qualitatively coincides with the
picture obtained in \cite{Arranz-98}. In both cases nonequidistant
levels appear in the same intervals of energy.
\begin{figure}[!h]
\centering
\includegraphics[width=8cm]{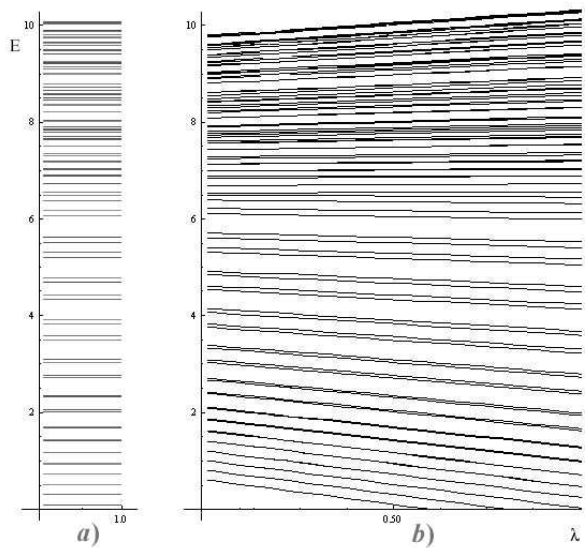}
\caption{The spectrum of lithium atom motion about the fragment
$C\equiv N$. Spectrum given in the Fig.5a is obtained from the
Fig.5b by means of projection on the energetic axis of the values
of energy at $\lambda=0.5$. \emph{a}) we see the nonequidistance
of energy levels of the lithium atom at energies
$\varepsilon\geq6\mbox{\ kcal/mol}$. In this region, the
transition from the oscillatory mode to the rotational one takes
place. \emph{b}) the energy spectrum depending on the parameter
$\lambda$.}\label{fig:5}
\end{figure}

\section{Conclusion}
\label{6}

Finding of energy spectrum of states which corresponds to the
process of isomerization comes to the finding of proper values of
Mathieu-Hill equation.

The energy spectrum of lithium atom motion includes a region
$\varepsilon<4 kcal/mol$, where levels are arranged equidistantly.
These levels correspond to small oscillations of the lithium atom
near isomeric states. Levels arranged nonequidistantly begin for
relatively high energies, when the energy level reaches half of
the barrier height and exceeds $\varepsilon\geq6 kcal/mol$.

It should be noted that the system considered in the paper is
integrable. Therefore the generality of nonequidistant (chaotic)
levels is not the manifestation of quantum chaos though it may
appear when these states are disturbed by periodic force as
illustrated by the example of a quantum pendulum
\cite{Ugulava-04,Ugulava-05}.

\parskip=\baselineskip
\textbf{Acknowledgement}

The financial support by the Deutsche Forschungsgemeinschaft (DFG)
through SFB 762, grant No. KO-2235/3 and STCU grant No 5053 is
gratefully acknowledged.



\end{document}